%% file: neurips_2026.tex
\documentclass{article}

\usepackage[position, final]{neurips_2026}

\usepackage[utf8]{inputenc} 
\usepackage[T1]{fontenc}    
\usepackage{hyperref}       
\usepackage{url}            
\usepackage{booktabs}       
\usepackage{amsfonts}       
\usepackage{nicefrac}       
\usepackage{microtype}      
\usepackage{xcolor}         
\usepackage{longtable}
\usepackage{longtable,booktabs,array}
\usepackage{calc} 
\usepackage{tabularx} 
\usepackage{graphicx} 

\patchcmd\longtable{\par}{\if@noskipsec\mbox{}\fi\par}{}{}
\IfFileExists{footnotehyper.sty}{\usepackage{footnotehyper}}{\usepackage{footnote}}
\makesavenoteenv{longtable}
\ifLuaTeX
  \usepackage{selnolig}  
\fi

\title{Workflow Closure Is Not Scientific Closure in Auto-Research Systems}

\author{%
    Shuai Wang\textsuperscript{1}, 
    Xinyuan Tian\textsuperscript{1}, 
    Pangpang Liu\textsuperscript{1}, 
    Yize Zhao\textsuperscript{1} \\  
    \textsuperscript{1}School of Public Health, Yale University, USA \\
    \texttt{\{shuai.wang.sw2572, yize.zhao\}@yale.edu}\\ 
}

\begin{document}
\makeatletter
\renewcommand{\@noticestring}{}
\makeatother

\maketitle

\begin{abstract}
   This paper argues that workflow closure is not scientific closure in auto-research systems. Current systems can increasingly complete research-like loops internally, moving from idea generation to experiment execution, writing, and self-evaluation. That achievement is real, but it does not by itself give the resulting outputs scientific standing. We argue that trustworthy auto-research should not aim for autonomous self-sufficiency, but should aim for autonomous execution under non-autonomous epistemic control. Based on a survey of more than 100 recent papers and repositories in this rapidly emerging area, together with a structured audit of 21 representative systems, we diagnose a recurring and structurally connected failure pattern: objective collapse, in which single-proxy targets replace multi-objective scientific aims; validation collapse, in which internal self-evaluation replaces independent validation; and acceptance collapse, in which benchmark scores or publication-shaped artifacts replace mechanisms for domain-level critique, reuse, and integration. These collapses are not inherent limits of autonomy but correctable design choices. Accordingly, we outline potential remedies across objective signal, validation, and output pathway to spark community discussion.
  
\end{abstract}

\input{./sections/section1.tex}
\input{./sections/section2.tex}
\input{./sections/section3.tex}
\input{./sections/section4.tex}
\input{./sections/section5.tex}
\input{./sections/section6.tex}
\input{./sections/section7.tex}
\input{./sections/section8.tex}
\input{./sections/section9.tex}
\input{./sections/section10.tex}

\newpage
\appendix
{\Large\bfseries APPENDIX}

\input{Appendix/Appendix_Survey}
\input{Appendix/Appendix_audits}

\newpage

\bibliographystyle{abbrv}
\bibliography{main}

\end{document}

%% file: sections/section1.tex
\section{Introduction}
\label{sec:introduction}

Auto-research has moved quickly from proposal to working systems. Karpathy's autoresearch release~\citep{andrej_karpathyautoresearch_2026} gave the community a spare, reusable template: propose a modification, run an experiment, inspect the outcome, and iterate inside a self-contained loop. \emph{Towards End-to-End Automation of AI Research}~\citep{lu_towards_2026}, published in \emph{Nature}, showed a more ambitious publication-shaped pipeline whose outputs passed workshop-level peer review at a major AI conference. Around these examples, auto-research has become a recognizable category of systems, repositories, and benchmarks~\citep{noauthor_sakanaaiai-scientist_2026, noauthor_aiming-labautoresearchclaw_2026, noauthor_hkudsai-researcher_2026, ferreira_can_2026, shen_empirical_2026, xu_asi_evolve_2026,noauthor_wecoaiaideml_2026,noauthor_samuelschmidgallagentlaboratory_2026}. The important question is no longer whether such systems can automate parts of a research workflow. It is what kind of closure they achieve when they do.

This work argues that workflow closure is not scientific closure in auto-research systems. Auto-research should not be evaluated as scientifically trustworthy merely because it closes an end-to-end research workflow; it should be designed so that objective plurality, independent validation, and domain-level critique enter the loop as architectural requirements.
 
The point is not that recent progress is illusory, or that any particular system must have overclaimed.
The point is that the field now has artifacts that make workflow closure look uncomfortably close to scientific closure.
The AI Scientist, for example, is described as moving from conception to publication and as having produced manuscripts that passed first-round workshop peer review~\citep{lu_towards_2026}; repository-centered systems such as \texttt{autoresearch} make internally repeated propose--execute--evaluate loops easy to instantiate~\citep{andrej_karpathyautoresearch_2026}.
These are important achievements, but they also make the central inference newly tempting: if a system can generate ideas, run experiments, write a paper, and evaluate its own output, then perhaps the research loop has been closed in the scientifically relevant sense.
This inference is the target of this paper.
A system achieves \emph{workflow closure} when it can execute an end-to-end research pipeline, from ideation through experimentation to writing and evaluation, without human intervention within a single cycle.
In such a system, the loop closes on itself: outputs feed back into subsequent decisions through the system's own evaluators.
By contrast, \emph{scientific closure} requires that research outputs remain answerable to evaluators, constraints, and uptake processes outside the system that produced them. 
Here, the loop must close through the world: through competing objectives rather than a single proxy, through independent validation rather than self-confirmation, and through mechanisms by which relevant communities can evaluate, contest, reuse, and integrate findings.
Workflow closure is therefore a real and important engineering achievement, but it does not by itself give the resulting outputs scientific standing.
 
Analyzing the recent wave of LLM-workflow auto-research systems, we introduce the concept \emph{three-level collapse} for the recurring pattern that arises when these systems are designed around closure-for-autonomy: making the loop close internally, cheaply, and repeatedly.
At the first level, \emph{objective collapse}, objective signal design reduces the plurality of scientific aims to a single optimizable proxy, making the loop ratchetable on a measure that no longer reliably tracks the property it was meant to indicate.
At the second level, \emph{validation collapse}, validator design keeps validators inside the producer's inductive boundary, such as self-review, same-family models, or benchmark-mediated signals, so that agreement among them cannot be assumed to constitute independence.
At the third level, \emph{acceptance collapse}, output pathway design terminates in scores or publication-shaped artifacts without standing pathways for the domain-level critique, reuse, and integration through which findings become usable knowledge.
These are not three independent problems but three faces of the same design choice.
The result is a drift toward local optimization: systems become increasingly good at improving what the loop can measure, validate, and present, while remaining insufficiently exposed to the external conditions required for scientific closure. However, these collapses are not inherent limits of autonomy but correctable design choices, and we propose potential remedies for each.

Against this background, this position paper contributes: 
(i) a distinction between workflow closure and scientific closure as architectural properties of the auto-research loop; 
(ii) a three-level collapse framework that diagnoses where contemporary systems fall short, grounded in a survey of more than 100 papers and repositories\ref{app:survey} and a structured audit of 21 representative systems \ref{app:audit}; 
(iii) design principles for auto-research systems that achieve autonomous execution under non-autonomous epistemic control.

%% file: sections/section2.tex
\section{Workflow Closure vs. Scientific Closure}
\label{sec:closure-definitions}
 
\textbf{Workflow closure} describes how a pipeline executes.
A system achieves workflow closure when it can execute an end-to-end research pipeline: covering ideation, experimentation, writing, and evaluation, without human intervention inside a single cycle, with evaluation signals sourced from within the system rather than from external validators. The key property is that the loop closes on itself: outputs generated at one stage feed back into subsequent system decisions through the system's own evaluators, scoring rules, or internal review mechanisms.
 
\textbf{Scientific closure} describes how its outputs answer to the epistemic operations required for knowledge production. Three conditions distinguish it from workflow closure: First, plural-objectives, where the output is evaluated against multiple non-reducible objectives rather than collapsed into a single signal. Second, independent-validation, where the output is checked by validators that are not functionally identical to the producer. Third, domain-uptake, where the research community can contest, reuse, and integrate the output into ongoing practice. Here the loop closes through the world: through constraints, evaluators, and pathways outside the system that produced it.

Achieving workflow closure does not imply scientific closure. Treating one as evidence of the other is a category error. Under closure-for-autonomy, each of the three conditions identified above is replaced by an internal substitute, producing what we call the three-level collapse: \textbf{Objective Collapse }into a single signal, \textbf{Validation Collapse} into self-checking, and \textbf{Acceptance Collapse} into terminal artifacts. Section ~\ref{sec:three-collapses} traces how each collapse arises, and how the three connect.

%% file: sections/section3.tex
\section{The Three-Level Collapse}
\label{sec:three-collapses}

\begin{figure*}[t]
    \centering
    \includegraphics[width=1\textwidth]{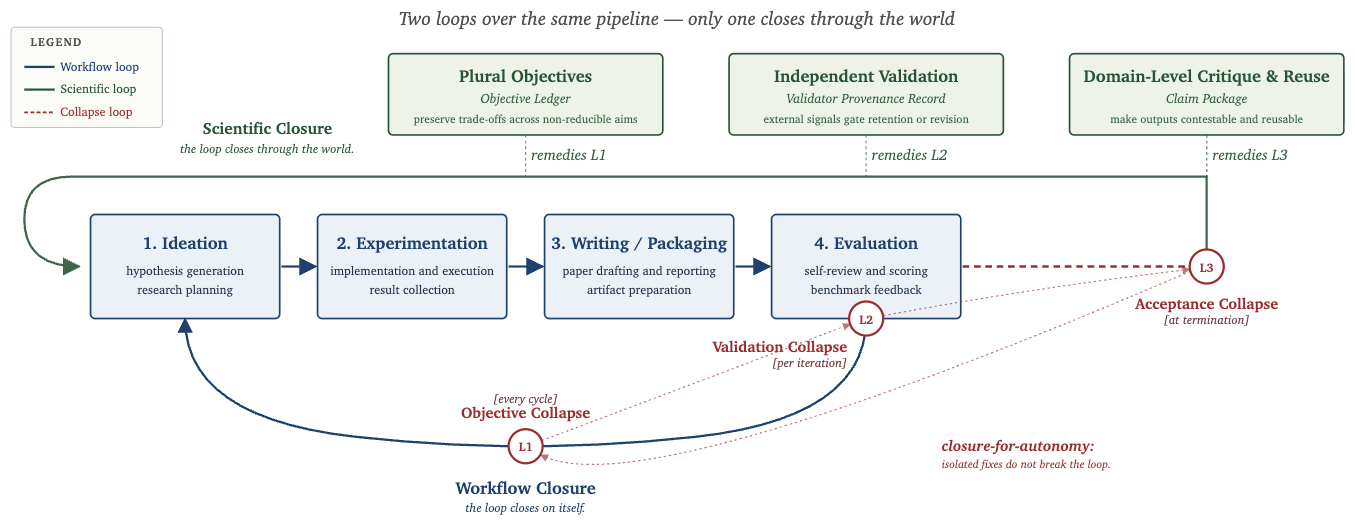}
    \caption{
    \textbf{Workflow Closure vs. Scientific Closure in Auto-Research Systems}
    }
    \label{fig:scientific-closure}
\end{figure*}

Once systems are optimized for closure-for-autonomy, the three conditions for scientific closure introduced in Section~\ref{sec:closure-definitions} are replaced by internal substitutes. Objective plurality is reduced to a single internal signal; independent validation is replaced by evaluation within the loop's own evaluative boundary; and domain-level uptake is replaced by termination in a score, report, or publication-shaped artifact. We call this pattern the three-level collapse: Objective Collapse, Validation Collapse, and Acceptance Collapse. These failures occur at three linked decision points in the loop: when the loop selects, when it validates, and when it terminates.

\subsection{L1: Objective Collapse}

\textbf{L1---objective collapse} occurs when progress under competing, non-reducible scientific objectives is reduced to a single internal signal. The relevant failure is not that a metric is used somewhere in the loop, but that retention, rejection, or ranking is governed by that metric alone. What Section~\ref{sec:closure-definitions} defined as objective plurality is thereby replaced by improvement under one comparable criterion.
 
This collapse follows directly from closure-for-autonomy. A loop designed to run internally, repeatedly, and at low coordination cost favors objective signals that are easy to compute, compare, and feed back into the next cycle. Plural objectives resist this pressure because they introduce trade-offs, incomparabilities, and occasions for external judgment. A single signal, by contrast, makes the loop ratchetable: modifications can be retained when the score improves and reverted when it does not. This design has clear engineering value, but it weakens the requirement that outputs remain robust under the objective plurality that scientific closure demands.

Objective collapse is a Goodhart-style failure~\citep{manheim2018categorizing,karwowski2023goodhart}: once a measure becomes the target of optimization, improvement on that measure no longer reliably tracks improvement in the broader property it was meant to indicate. Self-improving loops intensify the issue that the objective signal not only evaluates outputs but shapes the production of future ones, creating a stable pathway for over-optimization against the proxy. This is closely related to evidence that feedback loops with language models can induce in-context reward hacking~\citep{pan_feedback_2024}. In practice, L1 appears as monotonic improvement under the loop's internal score while the relationship between that score and the broader research objective becomes increasingly opaque.
 
\subsection{L2: Validation Collapse}

\textbf{L2---validation collapse} occurs when independent validation is replaced by internal evaluation, such as self-review, same-family evaluators, sibling-model checking, or benchmark-mediated signals that remain within the loop's own evaluative horizon. What Section~\ref{sec:closure-definitions} defined as independent validation is thereby replaced by success under internal validators. The relevant failure is not that internal evaluation occurs somewhere in the loop, but that validation is governed by internal evaluators alone.

Validation collapse is also encouraged by closure-for-autonomy. Independent validation is slow, costly, and difficult to standardize; it requires coordination with actors, distributions, or protocols that the loop does not itself control. An autonomous loop therefore has a strong incentive to internalize validation: a second model can score the first, an ensemble can simulate review, or a benchmark can serve as a standing gatekeeper. These choices preserve iteration speed and make the loop self-contained by removing the need for external validators.

The technical issue is not simply bias but correlated error induced by shared inductive structure. When producer and validator are trained, prompted, parameterized, or selected from closely related evaluative regimes, agreement between them cannot be assumed to constitute independent validation. The operational distinction that matters is whether an external signal actually enters the loop, rather than appearing after the fact as annotation, which is a distinction we return to in Section~\ref{sec:partial-remediation}. In practice, L2 appears as outputs that pass increasingly sophisticated review processes while remaining untested by validators outside the loop's own evaluative family.
 
\subsection{L3: Acceptance Collapse}
\textbf{L3---acceptance collapse} occurs when domain-level evaluation and integration are replaced by terminal artifacts, such as a score, a report, or a publication-shaped object that the loop produces without any mechanism for the research community to contest, reuse, or integrate it. What Section~\ref{sec:closure-definitions} defined as outputs subject to community evaluation thereby becomes outputs treated as if community evaluation had already occurred. 

Acceptance collapse is likewise encouraged by closure-for-autonomy. Domain-level evaluation and integration are costly because they depend on research communities, review processes, and field-specific evaluation standards that the system does not control. A loop designed to close internally has an incentive to stop earlier, at the point where the artifact can be scored, displayed, or packaged as complete, without being exposed to domain-level critique.

Acceptance collapse turns on a distinction between benchmark adequacy and domain standing. Benchmarks measure artifact quality under pre-specified tasks, but they do not instantiate the distributed process by which outputs become usable knowledge for a field. This process involves critique, reuse, and incorporation into ongoing practice, all of which are central to whether AI tools strengthen or weaken scientific understanding~\citep{messeri_artificial_2024}. Acceptance collapse is therefore not an objection to benchmarks themselves, but to the use of benchmark success as a substitute for domain-level evaluation. In practice, L3 appears as a growing set of outputs that appear complete because they are score-bearing, legible, and artifact-rich, while lacking systematic route into external critique or reuse.

\subsection{Why the Collapses Are Linked: Closure-for-Autonomy}

The three collapses are dynamically linked, forming a feedback loop stabilized by closure-for-autonomy. L1 enables L2 because a scalar objective can be checked internally, whereas plural objectives create pressure for external judgment. L2 enables L3 because, once internal validation is treated as sufficient, there is little architectural pressure to expose outputs to domain-level evaluation. L3 reinforces L1 because, once the endpoint is a score or artifact, the system is encouraged to optimize for the narrow criteria that make such endpoints look complete. The loop thus closes around what it can measure, validate, and terminate on its own.

This is why incremental fixes do not break the pattern. Adding a second metric is insufficient if validation stays internal. Adding an external checker is insufficient if its output does not gate the loop. Adding a stronger benchmark is insufficient if outputs still have no pathway into external critique or reuse. Each fix targets one collapse while leaving the other two intact, so the loop continues to close on itself. The three-level collapse is therefore the structural cost of closure-for-autonomy.

%% file: sections/section4.tex
\section{Collapse Patterns in Contemporary Auto-Research}
\label{sec:empirical-audit}

\input{tables/table_1}

Section~\ref{sec:three-collapses} introduced the \emph{three-level collapse} as a diagnostic framework. This section does not attempt to turn the position paper into a comprehensive benchmark study; rather, it uses a survey-backed audit to examine whether the proposed failure mode is visible in selected contemporary systems. In Appendix \ref{app:survey}, we systematically survey more than 100 recent papers and open-source repositories on LLM-workflow auto-research systems. From this corpus, we identify a full-coding audit pool of 21 systems with sufficient loop structure and public information for L1/L2/L3 coding (Table~\ref{tab:collapse-audit}). The scope is intentionally narrow: the audit includes LLM-workflow-based systems that implement an auto-research cycle, while excluding pre-LLM autonomous research systems, pure benchmarks, skill-only systems, and domain-specific deployment systems kept for contrast. Full methodology and per-system audit details are provided in Appendix~\ref{app:audit}.

Labels are assigned according to the role a design feature plays in the loop's control structure. For L1, a system is coded \textbf{strong} when retention, ranking, or continuation is governed by a single scalar, and \textbf{weak} when it broadens objective evidence but still resolves progress through internal aggregation. For L2, a system is coded \textbf{strong} when validation comes from self-review, same-family models, benchmark feedback, or other evaluators inside the producer's inductive boundary, and \textbf{weak} when it adds verifiers, cross-model review, citation checks, or adversarial probes that increase friction but do not gate the loop on independent validation. For L3, a system is coded \textbf{strong} when the terminal endpoint is a score, report, or publication-shaped output without a standing pathway by which the research community can evaluate, contest, or reuse it, and \textbf{weak} when external contact exists but is post-hoc or does not feed back into the loop. The label \emph{mitigated} is reserved for designs that fully resolve the corresponding collapse, which no LLM-workflow auto-research system in the surveyed corpus achieves. 
 
The audit's most informative finding is the asymmetry across the three collapse dimensions. Among the audited systems, L1-strong is 81.0\% (17/21), L2-strong is 71.4\% (15/21), and L3-strong is 90.5\% (19/21). L3 is the least remediated layer: only 9.5\% (2/21) of systems fall into the weak category, suggesting that domain-level evaluation and integration remain the weakest part of the current paradigm. L2 is the most remediated layer: 28.6\% (6/21) are L2-weak, indicating that current systems most often try to move beyond fully internal closure through validation-side checks. L1 sits between them: 19.0\% (4/21) are L1-weak, typically through mechanisms that broaden the information considered before selection, such as reflective critique, multi-objective aggregation, or parallel search.

The broader corpus also contains contrast cases that clarify the boundary of the diagnosis. Four systems exhibit at least one mitigated label because an external operation is architecturally required before the loop can treat an output as valid. In the semi-autonomous formalization of the VML equilibrium, proof generation is coupled to Lean 4 verification and in-loop mathematician supervision, mitigating L1 and L2 because objective validity is fixed by formal proof correctness and validation is not reducible to internal self-review~\citep{ilin_semi-autonomous_2026}. In Latent-Y, antibody design campaigns are integrated with wet-lab execution, so L2 is mitigated by a reality-grounded validator rather than an internally accessible judge~\citep{team_latent-y_2026}. Robin extends this laboratory-grounded pattern to autonomous therapeutic 
discovery: although built on multi-agent literature search and data 
analysis that resemble closure-for-autonomy systems on the surface, its architectural core is the in-vitro experimental loop in which candidate molecules selected by the agent are confirmed by laboratory assays before the loop treats them as validated~\citep{Ghareeb2026-sk}. CRISPR-GPT provides a fourth route: expert sign-off is architecturally required at critical junctures, making external human validation part of the loop rather than a post-hoc annotation~\citep{noauthor_crispr-gpt_2024}. These cases sit outside the main LLM-workflow closure-for-autonomy paradigm, but that is precisely why they are informative. They support a bounded claim: the three-level collapse is not an intrinsic cost of autonomous research as such, but a consequence of a specific contemporary design philosophy.

The audit supports two bounded conclusions. First, among the audited systems, the three-level collapse is informatively asymmetric: L3 remains the least remediated dimension, L2 is where partial remediation is most often attempted, and L1 sits between them. Second, contrast cases outside the internally closed closure-for-autonomy paradigm show that at least one collapse dimension can be mitigated when external operations become part of the loop's validity conditions. The audit does not show that the pattern is universal across all autonomous research, nor that the observed co-occurrence proves causation, nor that the audit pool is exhaustive. The weak labels in Table~\ref{tab:collapse-audit} mark departures from strong collapse within the same paradigm; Section~\ref{sec:partial-remediation} examines these departures together with related remediation attempts in the broader corpus.

%% file: tables/table_1.tex
\begin{table*}[t]
\centering
\caption{Collapse audit of 21 auto-research systems in the full-coding audit pool. \textbf{S} = strong collapse; \textbf{W} = weak collapse (remediation attempted but not architecturally sufficient). }
\label{tab:collapse-audit}
\small
\setlength{\tabcolsep}{4pt}
\renewcommand{\arraystretch}{1.05}
\begin{tabularx}{\textwidth}{@{}X c c c X@{}}
\toprule
System & L1 & L2 & L3 & Notes \\
\midrule
Karpathy autoresearch~\citep{andrej_karpathyautoresearch_2026} & S & S & S & Canonical single-scalar loop \\
uditgoenka/autoresearch~\citep{noauthor_uditgoenkaautoresearch_2026} & S & S & S & Directly follows autoresearch loop \\
goal-md~\citep{noauthor_jmilinovichgoal-md_2026} & S & S & S & Agent-constructed specialized metric \\
autoresearch-anything~\citep{noauthor_zkarimi22autoresearch-anything_2026} & S & S & S & Task-specific metric\\
ADAS~\citep{noauthor_shengranhuadas_2026} & S & S & S & Meta-optimization on agent architecture \\
AIDE~\citep{noauthor_wecoaiaideml_2026} & S & S & S & Benchmark-driven and tree search \\
GEPA~\citep{noauthor_gepa-aigepa_2026} & W & S & S & Reflection adds minor L1 nuance \\
HGM~\citep{noauthor_metauto-aihgm_2026} & S & S & S & Self-improvement on SWE-bench \\
Bilevel Autoresearch~\citep{qu_bilevel_2026} & S & S & S & Loop-on-loop meta-optimization \\
ASI-Evolve~\citep{xu_asi_evolve_2026} & W & S & S & Three parallel objectives, scalar-aggregated \\
Omni-SimpleMem~\citep{noauthor_omni-simplemem_2026} & S & S & S & Autoresearch on agent memory \\
Claudini~\citep{panfilov_claudini_2026} & S & W & S & Held-out adversarial checks \\
CORAL~\citep{noauthor_coral_2026} & S & W & S & Evaluator separation, shared memory \\
SakanaAI/AI-Scientist~\citep{noauthor_sakanaaiai-scientist_2026,lu_towards_2026} & S & W & W & Real workshop review \\
AI-Scientist-v2~\citep{noauthor_sakanaaiai-scientist-v2_2026} & S & S & S & Extension of AI-Scientist pipeline \\
AutoResearchClaw~\citep{noauthor_aiming-labautoresearchclaw_2026} & S & W & S & Narrow external check (citation verification) \\
Agent Laboratory~\citep{noauthor_samuelschmidgallagentlaboratory_2026,schmidgall_agentrxiv_2025} & S & S & S & Three-phase pipeline; AgentRxiv-based sharing \\
CycleResearcher~\citep{wengcycleresearcher} & S & S & S & Paired policy/reward LLMs, focused on writing \\
AI-Researcher~\citep{noauthor_hkudsai-researcher_2026} & W & W & W & Production development with external contact \\
ARIS~\citep{noauthor_wanshuiyinauto-claude-code-research--sleep_2026} & S & W & S & Cross-model review, skills-based \\
ClawTeam~\citep{noauthor_hkudsclawteam_2026} & W & S & S & Swarm, parallel GPU directions \\
\bottomrule
\end{tabularx}
\end{table*}

%% file: sections/section5.tex
\section{How Current Systems Attempt to Remediate the Collapse}
\label{sec:partial-remediation}

Multiple recent systems attempt partial remediation of the three-level collapse. In the closure-for-autonomy framework of Section~\ref{sec:three-collapses}, these attempts share two recurring limits. First, they are \emph{not systematized}: added checks are usually ad hoc additions rather than pre-defined discriminative protocols. Second, they remain \emph{post-hoc rather than in-loop}: external signals rarely become correction signals inside the system's primary research loop.

At the level of objective design, several systems broaden the internal evidence available to the loop before a decision is made. ASI-Evolve distributes search across neural architecture design, data curation, and reinforcement-learning algorithm discovery, demonstrating that auto-research can coordinate multiple research fronts within one system~\citep{xu_asi_evolve_2026}. GEPA introduces reflective prompt evolution, using natural-language analysis of failed mutations to shape subsequent ones, which turns mutation search into a more interpretable process than pure scalar ratcheting~\citep{noauthor_gepa-aigepa_2026}. ClawTeam expands exploration through swarm-style parallel GPU directions, increasing breadth relative to serial loops~\citep{noauthor_hkudsclawteam_2026}. goal-md extends the paradigm further by having the agent construct its own fitness function from a natural-language goal before optimization begins~\citep{noauthor_jmilinovichgoal-md_2026}. These remedies enrich what the loop considers but not how it decides: a scalar aggregation still governs retention, ranking, or selection. They remain non-systematized, with each system broadening objective evidence in its own way, and partial rather than structural: the added evidence may enter the loop, but it does not become a non-reducible objective constraint or an external validity condition.

At the level of validation, the engineering moves are even more substantial. Marco DeepResearch makes verification a first-class pipeline component, exposing the verifier as inspectable infrastructure rather than burying judgment inside the generator~\citep{zhu_marco_2026}. MiroThinker-H1 integrates adversarial probing into the mid-training stage, so verification pressure is applied before final outputs are frozen~\citep{team_mirothinker-17_2026}. ARIS introduces cross-model review loops, increasing validator diversity beyond single-model self-review~\citep{noauthor_wanshuiyinauto-claude-code-research--sleep_2026}. AutoResearchClaw adds citation verification and multi-batch coordination, bringing a narrow but genuine external check into the loop~\citep{noauthor_aiming-labautoresearchclaw_2026}. These attempts are non-systematized because each system defines validation differently and none establishes a general protocol for distinguishing substantive correctness from internally legible agreement. They also remain partial rather than structural: citation checks verify references rather than claims, cross-model review stays within the LLM family, and adversarial probes remain self-generated or system-internal.
 
At the level of output pathways, remediation attempts are rarer and more externally meaningful. AI-Researcher provides a rare data point on production-facing deployment through novix.science, moving beyond benchmark-only evaluation~\citep{noauthor_hkudsai-researcher_2026}. SakanaAI/AI-Scientist submitted generated papers to real workshop peer review, testing publication-shaped outputs against human expert judgment rather than only internal evaluators~\citep{noauthor_sakanaaiai-scientist_2026,lu_towards_2026}. Their common architectural limit is that the engagement remains non-systematized and post-hoc rather than in-loop. Production use is evidence of contact with the world, not yet a documented mechanism by which user feedback drives the loop's evolution; a workshop submission is an event, not a standing pathway for contestation, reuse, and integration.

The architectural lesson is that partial remediation on one dimension is readily reabsorbed by the closure structure of the other two. Breaking the collapse requires coordinated remediation across all three design choices: objective signal, validator, and output pathway. Among the audited systems, no system attempts that coordination, and Section~\ref{sec:constructive-agenda} turns to its constructive implications.

%% file: sections/section6.tex
\section{Toward Scientific Closure: Design Implications for Auto-Research Systems}
\label{sec:constructive-agenda}
 
The diagnosis in Sections~\ref{sec:three-collapses}--\ref{sec:partial-remediation} has direct design implications. The relevant question is not which single architecture would solve the problem, but which architectural commitments would move auto-research from internal loop completion toward scientific closure. The position is not that auto-research should be slowed or made to resemble traditional human research. It is that auto-research should preserve autonomous execution while giving up epistemic self-sufficiency. The design target therefore shifts from closure on itself toward closure through the world.

Against objective collapse, \emph{objective signal design} should treat plural objectives as architectural primitives rather than as inputs to be aggregated away at the moment of decision. Where appropriate, evaluation should operate over a Pareto frontier of non-reducible objectives, so that trade-offs remain visible to the loop and retention decisions preserve them rather than prematurely reduce them to a single notion of progress. Concretely, auto-research systems should maintain an \emph{objective ledger}: a structured record of the aims a candidate output is supposed to satisfy, the evidence for each, the trade-offs among them, and the decision rule used when the loop proceeds despite unresolved conflict. For a machine-learning method paper, such a ledger might separate predictive performance, calibration, robustness, computational cost, interpretability, and reproducibility. For a biomedical discovery paper, it might separate mechanistic plausibility, assay feasibility, safety constraints, and clinical relevance. The point is not that every domain shares the same objective set, but that the loop represents objective plurality explicitly enough that competing criteria remain part of the control structure. 

Against validation collapse, \emph{validator design} should make external validation both architecturally required and in-loop: a validator's signal should enter the research loop as a condition of validity, not annotate outputs after the loop has already decided what counts as progress. Concretely, systems should maintain a \emph{validator provenance record}: who or what validated each claim, the validator's independence relation to the producer, the evidence inspected, and whether the result changed subsequent search. A same-family LLM reviewer may be useful for debugging, but it should not be counted as independent validation unless its signal is tied to an external constraint. Depending on the domain, the relevant validator could be a formal proof checker, a pre-locked holdout benchmark, a wet-lab assay, an external replication cohort, a human expert panel, or a protocol that forces claims through independently maintained tools and data. What matters is not the validator type, but that the validator lies outside the producer's inductive boundary and that its signal materially alters whether outputs are retained, revised, or rejected.

Against acceptance collapse, \emph{output pathway design} should create architectural pathways through which outputs can enter domain-level evaluation, critique, reuse, and integration. The requirement is weaker than formal peer review or institutional approval: what matters is not whether a specific bureaucratic channel is traversed, but whether outputs become contestable and reusable outside the producing loop. Concretely, systems should emit a \emph{claim package} rather than only a paper-shaped artifact: a set of atomic claims, their evidence traces, code and data provenance, failure cases, known limitations and sensitivities, and a route by which external users can reproduce, contest, or extend the claims. For domains with high downstream stakes, that package should also include stopping rules: conditions under which the system must defer, request external review, or decline to treat an output as ready for domain uptake. The design question is not how to assign a stronger terminal label to an artifact, but how to ensure that an artifact's standing can be tested outside the loop. In this sense, \emph{acceptance is a process, not a score}. 

These three directions imply a single meta-principle: \emph{autonomous execution under non-autonomous epistemic control}. Auto-research should move from loops that close on themselves to loops that close through external objectives, external validators, and external evaluation pathways. Autonomy as execution capability would be preserved; autonomy as epistemic self-sufficiency would be abandoned. Trustworthy auto-research then becomes a tractable architectural problem rather than an open-ended ideal: for each design choice, the question is whether the loop remains self-contained, or whether it closes through the world.

%% file: sections/section7.tex
\section{Related Work and Positioning}
\label{sec:related-work}
This section positions the paper against the two bodies of work most directly tied to the main argument: LLM-workflow auto-research systems, which establish the target phenomenon, and concurrent diagnostic positions, which diagnose adjacent failures at the level of behavior or epistemics rather than loop architecture. A more comprehensive survey of recent LLM-workflow auto-research is provided in Appendix~\ref{app:survey}.

LLM-workflow auto-research systems establish the target phenomenon of this paper. Recent surveys describe a coherent technical agenda in which proposal, execution, evaluation, and revision are composed into internally iterating research loops \citep{wei_ai_2025}, with the broader categorical landscape situating autonomous scientific discovery alongside hybrid human-AI co-creation as identified phases of FM-driven scientific research \citep{liu2025foundation}. A complementary lifecycle-stage view organizes the same field by epistemological phase rather than by loop architecture, mapping AI assistance across creation, writing, validation, and dissemination~\citep{Kong2026-as}. Concrete instantiations range from minimal repository templates \citep{andrej_karpathyautoresearch_2026,noauthor_uditgoenkaautoresearch_2026} to multi-stage publication-shaped pipelines \citep{lu_towards_2026,schmidgall_agentrxiv_2025} and meta-level designs that optimize the loop itself \citep{qu_bilevel_2026}. These systems differ in domain, ambition, and degree of automation, but they share a control structure in which candidate outputs are generated, scored, retained, revised, and packaged through signals available inside the loop. The argument of this paper is therefore not about the quality or limitations of any single system. It is about what happens when this shared loop structure becomes the basis for deciding progress, validity, and standing.

Concurrent positions are complementary, but they operate at a
different level. Some locate failure in agent reasoning behavior, including weak evidence consideration, limited refutation-driven revision, and lack of convergent multi-test reasoning \citep{rios-garcia_ai_2026}. Others show empirically that nominally broad architecture search can reduce to narrow tuning within a single design region \citep{li_auto_2026}, or that agent-generated papers that score competitively under manuscript-only review fall short under artifact-aware review, with experimental rigor as
the dominant failure mode~\citep{Zhang2026-fg}. Editorial and
analytical commentary in \emph{Nature} further argues that AI systems can empower scientific work but should not replace human
researchers~\citep{noauthor_2026-lw}, and that AI tools can produce illusions of understanding when they bypass the distributed processes through which scientific knowledge is critiqued, integrated, and revised \citep{messeri_artificial_2024}. Together these works ask whether individual runs reason well, whether nominal search is genuinely broad, whether artifact quality survives artifact-aware scrutiny, and whether artifact production substitutes for knowledge production. Our diagnosis asks a different and compatible question: how the surrounding control structure decides that progress, validity, and standing have been achieved. 

%% file: sections/section8.tex
\section{Limitations}
\label{sec:limitations}
 
This paper's scope is LLM-workflow auto-research, and the diagnosis may not transfer unchanged to formal-verification, robotic-experiment, or other autonomous-research paradigms. The audit covers selected systems with sufficient loop structure and public information for full L1/L2/L3 coding, rather than the full population, so the reported rates should not be read as prevalence estimates across all uses of LLMs in science. The three-level collapse is the most common and structurally connected failure pattern we identify in this category, not an exhaustive account of all possible failures.

%% file: sections/section9.tex
\section{Alternative Views and Scope Boundaries}
\label{sec:alternative-views}

This section addresses five objections most directly tied to the paper's central claim.

A first objection is that workflow closure is only an engineering milestone, not a claim of scientific standing. We agree that many system builders are appropriately cautious, and our argument does not depend on attributing overclaiming to any individual work. The issue is structural: once systems are benchmarked, advertised, and compared by their ability to move from idea generation to paper-shaped output, workflow completion can become the most visible proxy for research success. This paper calls on the community to resist that proxy before it becomes naturalized.

A second objection is that benchmarks and internal validators are necessary because external validation does not scale. This is true, but it does not answer the concern. Benchmarks are indispensable for development, debugging, and comparison; internal validators are indispensable for rapid iteration. The problem arises when these instruments are treated as substitutes for the external operations they can only approximate. The constructive agenda in Section~\ref{sec:constructive-agenda} therefore does not ask every iteration to wait for full peer review or wet-lab validation. It asks that systems represent the difference between internal progress signals and external validity conditions, and that the latter enter the loop at decision points where claims acquire standing.

A third objection is that diverse model ensembles, adversarial agents, or specialized verifier models may provide independence without human or domain-level validation. Such methods can reduce correlated error and should be part of the design space. But independence is not a property of multiplicity alone: models trained on overlapping data, selected by the same benchmark culture, or rewarded under the same scoring rules can still share the producer's blind spots. The relevant question is whether the validator introduces an external constraint that the producing loop cannot freely simulate, optimize against, or reinterpret as success. Artificial validators may eventually satisfy that condition, but their independence must be demonstrated rather than assumed.

A fourth objection is that science has always relied on proxies, publication artifacts, and imperfect peer review, so auto-research is not categorically different. We agree that the problem is not unique to auto-research, and much human-authored machine-learning work would also fail a demanding L3 criterion. What is distinctive is the speed and closure of the feedback loop. Human scientific communities contain partially independent frictions across laboratories, methods, incentives, instruments, and communities; auto-research systems can compress ideation, execution, validation, and writing into a single internally optimized loop. That compression is the source of their power, and also why external constraint must be designed architecturally.

A fifth objection is that canonical systems such as Coscientist, FunSearch, AlphaEvolve, Google AI Co-Scientist, and OpenAI Deep Research falsify the universality claim. This objection clarifies rather than refutes the paper's scope. The claim is not that every LLM-assisted scientific system exhibits the three-level collapse, but that collapse is characteristic of systems whose research loop closes primarily through internally available objectives, validators, and terminal artifacts. Coscientist, FunSearch, and AlphaEvolve differ because external, executable, or production-relevant constraints are constitutive parts of the architecture: Coscientist incorporates cloud-lab execution and chemical experimentation, FunSearch filters LLM-generated programs through executable evaluators, and AlphaEvolve depends on executable algorithmic evaluation and, in some settings, production-relevant computational constraints~\citep{boiko_autonomous_2023,romera-paredes_mathematical_2023,novikov_alphaevolve_2025}. The same architectural pattern is visible in two recent systems published on \emph{Nature} that further sharpen the scope line, one on the executable-evaluation axis and one on the laboratory-grounded axis. ERA couples LLM-driven code generation with tree search over executable quality metrics on public leaderboards across single-cell analysis, COVID-19 hospitalization forecasting, and additional scientific software tasks, so that retention is
gated by external execution rather than by internal evaluators
alone~\citep{Aygun2026-qy}. Robin proceeds further along the
laboratory-grounded axis: it integrates multi-agent literature search and
data analysis with an in-vitro experimental loop, autonomously identifying
ripasudil and KL001 as therapeutic candidates for dry age-related macular
degeneration whose efficacy is then confirmed in the
laboratory~\citep{Ghareeb2026-sk}. In this design, lab assays are not
post-hoc validation of a completed paper but constitutive validity
conditions on each design cycle, in the same architectural sense as
Coscientist. Google AI Co-Scientist and OpenAI Deep Research differ in a second way: the
former is primarily a scientist-guided hypothesis-generation system whose
outputs require downstream experimental validation, with its \emph{Nature}
2026 deployment reporting in-vitro confirmation of drug repurposing
candidates for acute myeloid leukemia~\citep{gottweis_towards_2025}, while
the latter is a source-grounded research-synthesis product rather than an
autonomous discovery loop~\citep{openai_deep_research_2025}. These systems
therefore support the paper's scope line rather than undermine it.

%% file: sections/section10.tex
\section{Conclusion}
\label{sec:conclusion}

Auto-research is increasingly at risk of being treated as if workflow closure were sufficient evidence of scientific closure, yet the inference from one to the other is structurally unsound. In this position paper, we identify the core risk as a three-level collapse: objective collapse, validation collapse, and acceptance collapse. We further derive design implications for objective signal design, validator design, and output pathway design, arguing that these collapses are correctable design choices rather than inherent limits of autonomy. The path forward is to preserve autonomous execution while giving up epistemic self-sufficiency: auto-research systems should be designed not to close on themselves, but to close through the world.

%% file: Appendix/Appendix_Survey.tex
\section{A Survey of LLM-Workflow Auto-Research}\label{app:survey}

\subsection{Scope and Methodology of the Survey}\label{app:survey:scope}

This appendix surveys the landscape of LLM-workflow auto-research as of early 2026. By LLM-workflow auto-research, we mean systems in which language-model agents plan, generate, evaluate, revise, or package research artifacts through a multi-stage workflow, executing more of the research process than question answering or single-step generation typically covers. The corpus surveyed here corresponds to the inventory described in the main text's audit methodology, more than one hundred works spanning archival papers and open-source repositories, with the time window concentrated in 2025--2026 but including foundational earlier work such as Coscientist~\citep{boiko_autonomous_2023} and FunSearch~\citep{romera-paredes_mathematical_2023}. Several categories are intentionally excluded: pure literature-retrieval systems labeled as ``research'' that do not iterate, generic coding agents not framed for research automation, traditional AutoML and hyperparameter optimization systems that predate the LLM-workflow paradigm, and general LLM-agent frameworks without a documented research-automation instantiation. Adjacent systems with constitutive external operations, such as formal verification, wet-lab integration, executable evaluation, are retained as boundary cases in Appendix~\ref{app:survey:boundary}.

The remainder of this appendix is organized along five lines. Appendix~\ref{app:survey:ecosystem} covers the core auto-research loop ecosystem; Appendix~\ref{app:survey:substrates} covers the substrate capabilities, especially skill and memory systems, that auto-research loops increasingly compose over; Appendix~\ref{app:survey:domains} covers domain-specific applications and boundary systems; and Appendix~\ref{app:survey:concurrent} discusses concurrent diagnostic positions on auto-research as a category. 

The survey provides a timely and objective overview of this rapidly evolving field, while the position paper highlights perspectives and recommendations that deserve urgent attention. The two parts support each other.

\subsection{The Auto-Research Loop Ecosystem}\label{app:survey:ecosystem}

This paper organizes the auto-research ecosystem into four subcategories that together cover the main design archetypes currently being explored. \emph{Foundational templates} establish the minimal loop structure and seed a broad ecosystem of forks and adaptations. \emph{Publication-shaped end-to-end pipelines} extend the loop across the full research workflow from ideation to manuscript drafting. \emph{Meta-level and self-improving systems} treat the search procedure or the agent itself as the object being improved. \emph{Multi-agent collaborative systems} distribute the loop across coordinated agents. The categories are not mutually exclusive, but together they capture the dominant designs in the current literature.

\subsubsection{Foundational Templates and Direct Descendants}\label{app:survey:foundational}

The contemporary auto-research ecosystem traces a clear lineage to Karpathy's autoresearch repository \citep{andrej_karpathyautoresearch_2026}, which distills the auto-research loop into a minimal four-step pattern: modify code, verify against a benchmark, keep or discard based on the result, and repeat. The design is deliberately spare. A single LLM agent operates over training code on a single GPU, with no orchestration layer, no multi-agent coordination, and no external validators. This minimalism is what made the template generative: the entire loop fits in a few hundred lines, the modification--verification cycle is easy to reason about, and the closure-for-autonomy design pattern is exposed in its simplest form, where outputs at each step feeding back into the loop through internally computed scores. The template's influence is visible both in repositories that fork it directly and in research that uses it as a controlled testbed.

A first cluster of descendants ports the template across LLM platforms while preserving its core control structure. uditgoenka/autoresearch\citep{noauthor_uditgoenkaautoresearch_2026} adapts the loop to Claude Code; supratikpm/gemini-autoresearch\citep{noauthor_supratikpmgemini-autoresearch_2026} targets the Gemini CLI; leo-lilinxiao/codex-autoresearch \citep{noauthor_leo-lilinxiaocodex-autoresearch_2026} brings the same pattern to Codex; and davebcn87/pi-autoresearch \citep{noauthor_davebcn87pi-autoresearch_2026} together with drivelineresearch/autoresearch-claude-code\citep{noauthor_drivelineresearchautoresearch-claude-code_2026} extend the loop to additional agent runtimes. A second cluster generalizes the template beyond its original ML-training setting. jmilinovich/goal-md \citep{noauthor_jmilinovichgoal-md_2026} replaces the fixed benchmark with an agent-constructed fitness function specified in a natural-language goal file, and zkarimi22/autoresearch-anything \citep{noauthor_zkarimi22autoresearch-anything_2026} parameterizes the loop over arbitrary task definitions. Cross-platform packaging efforts including Entrpi/autoresearch-everywhere \citep{noauthor_entrpiautoresearch-everywhere_2026}, mutable-state-inc/autoresearch-at-home \citep{noauthor_mutable-state-incautoresearch-at-home_2026}, eimenhmdt/autoresearcher \citep{noauthor_eimenhmdtautoresearcher_2026}, greyhaven-ai/autocontext \citep{noauthor_greyhaven-aiautocontext_2026}, and james-s-tayler/lazy-developer \citep{noauthor_james-s-taylerlazy-developer_2026} further illustrate how rapidly the template propagated as a reusable substrate. He et al.\ \citep{he_autoresearch_2026} use the same template as the entry point for a methodological position statement on how the human role shifts from experimenter to research director, a perspective we return to in \ref{app:survey:concurrent}.

The template has also served as a testbed for controlled studies of what auto-research loops actually do. Ferreira et al.\ \citep{ferreira_can_2026} compare classical hyperparameter-optimization algorithms against LLM-based search inside the auto-research framework, fixing the compute budget and varying only the optimizer. Within a fixed search space, classical methods such as CMA-ES and TPE consistently outperform LLM-based agents; however, when the LLM agent is allowed to edit training source code in an unconstrained space, the gap narrows substantially even with a self-hosted open-weight model. The authors further find that reliability dominates exploration breadth: methods that avoid out-of-memory failures outperform those with higher search diversity. Their experiments suggest that the auto-research loop benefits less from scaling the agent than from coupling it with a structured external optimizer. A complementary diagnostic study by Li \citep{li_auto_2026}, using the same template at 10{,}000-experiment scale, examines whether the loop's nominal architecture search collapses into hyperparameter tuning within a narrow design region; we discuss its implications in \ref{app:survey:concurrent}.

\subsubsection{Publication-Shaped End-to-End Pipelines}\label{app:survey:publication-shaped}

A second branch of the auto-research ecosystem aims for substantially broader scope than the foundational templates: rather than iterating over a single training script, these systems attempt to cover the full research pipeline from ideation through experimentation, writing, and self-review, producing outputs in publication-shaped forms. Lu et al.'s AI Scientist~\citep{lu_towards_2026} is the most visible exemplar. Published in \textit{Nature}, the system composes literature-grounded idea generation, experimental code synthesis, automated execution, and LaTeX paper drafting into a single end-to-end loop, with an internal LLM reviewer scoring the final manuscript. Several generated manuscripts were submitted to a workshop at a major machine-learning venue and received first-round peer-review acceptance scores. The system has set the architectural template that much of the subsequent publication-shaped category adapts or extends.

Several systems extend the paradigm with collaborative or production-facing elements. Schmidgall and Moor's AgentLaboratory~\citep{noauthor_samuelschmidgallagentlaboratory_2026,schmidgall_agentrxiv_2025} structures the pipeline into three phases (literature review, experimentation, writing) and introduces AgentRxiv, a shared repository through which agents can access papers produced by other agent instances, introducing a form of inter-agent knowledge transfer. HKUDS AI-Researcher~\citep{noauthor_hkudsai-researcher_2026} takes a different path: it pairs the auto-research loop with a production-facing deployment at novix.science, exposing the system to external users beyond benchmark evaluation. SakanaAI/AI-Scientist-v2~\citep{noauthor_sakanaaiai-scientist-v2_2026} continues the original AI Scientist line with refinements to the experimental and writing modules. Adjacent systems including AutoResearchClaw~\citep{noauthor_aiming-labautoresearchclaw_2026,Liu2026-az}, which adds citation verification and multi-batch coordination, BloClaw~\citep{qin_bloclaw_2026}, a multimodal scientific workspace, and the Sibyl auto-research system~\citep{noauthor_sibyl-research-teamautoresearch-sibylsystem_2026} further illustrate the architectural diversity within the paradigm.

A further group situates the publication-shaped paradigm in distinct methodological or evaluative regimes. CycleResearcher~\citep{wengcycleresearcher} trains paired policy and reward LLMs through SimPO using historical ICLR review data, producing a system in which the writing process is optimized against learned reviewer preferences. Earlier systems such as MLR-Copilot~\citep{noauthor_du-nlp-labmlr-copilot_2025} and ResearchAgent~\citep{noauthor_jinheonbaekresearchagent_2026} established proto-versions of the pipeline with stronger human-in-the-loop elements at ideation and revision. Lighter-weight implementations including NanoResearch~\citep{noauthor_openraisernanoresearch_2026} and OpenAGS~\citep{noauthor_openagsopenags_2026} demonstrate the template's portability to smaller computational budgets, while AutoSOTA~\citep{noauthor_autosota_2026} narrows the scope to automated discovery of state-of-the-art configurations within specified model families.

Beyond the publication-shaped pipelines themselves, several literature-handling components are frequently composed into auto-research loops at the literature-retrieval and ideation stages. LitLLM~\citep{noauthor_litllmlitllm_2026} provides an open framework for literature-aware LLM research workflows; LatteReview~\citep{noauthor_pouriarouzrokhlattereview_2026} formulates agent-driven literature review as a structured retrieval-and-synthesis task; and AwesomeLit~\citep{xie_awesomelit_2026} extends this line toward hypothesis generation, with agent-supported literature research feeding into the formulation of new research questions. These components do not themselves close an auto-research loop, but they supply the literature-grounding substrate that publication-shaped pipelines depend on.

\subsubsection{Meta-Level and Self-Improving Systems}\label{app:survey:meta-level}

A third branch of the auto-research ecosystem moves up a level: rather than executing research on a fixed task, these systems treat the search procedure itself or the agent that performs the search as the object being improved. ASI-Evolve~\citep{xu_asi_evolve_2026} distributes autonomous search across three parallel research frontiers, including neural architecture design, training-data curation, and reinforcement-learning algorithm discovery, which is demonstrating that an auto-research system can coordinate substantive research directions across multiple sub-fields within a single coordinated framework, with cross-front signals aggregated into a system-level performance measure. Bilevel Autoresearch~\citep{qu_bilevel_2026} takes the meta-move in a different direction: it instantiates an outer loop whose object of optimization is the inner research loop itself, with the outer loop proposing modifications to the inner loop's search procedure based on how rapidly the inner loop improves on its task. The two systems together mark out the design space for meta-level auto-research: ASI-Evolve expands the breadth of \emph{what} the loop searches, while Bilevel Autoresearch lifts the level of \emph{how} the search is conducted.

A complementary line targets self-improvement at fixed evaluative endpoints. HGM~\citep{noauthor_metauto-aihgm_2026} formulates self-improvement against SWE-bench, with the agent revising its own codebase, tooling, and prompt structure across iterations as SWE-bench performance changes. The system illustrates the general pattern: a benchmark provides the optimization signal, and the agent's own configuration becomes the modification surface. Claudini~\citep{panfilov_claudini_2026} applies the same logic in a substantively different domain, using an autoresearch-style loop to discover novel adversarial-attack algorithms against LLMs. Distinct from most systems in this category, Claudini supplements the internal search loop with a held-out adversarial evaluation that the loop cannot optimize against directly, demonstrating how the self-improving template can be combined with a partial external check while remaining structurally loop-driven. GRAFT-ATHENA~\citep{Toscano2026-td} extends self-improvement to evolutionary numerical-algorithm discovery, in which agentic teams iteratively refine numerical solvers against fixed quantitative performance criteria.

Adjacent designs explore other surfaces on which the loop's own machinery can be modified. ADAS~\citep{noauthor_shengranhuadas_2026} treats the multi-agent system architecture itself as the search space, with one agent proposing variants of agentic structure that are evaluated on downstream benchmark tasks. AIRA~\citep{Pepe2026-tc} pursues an analogous move at the neural-architecture level, with compose-and-design agents searching over foundation-model architectures beyond standard Transformers. GEPA~\citep{noauthor_gepa-aigepa_2026} introduces reflective prompt evolution, using natural-language analysis of failed mutations to guide subsequent prompt modifications and yielding a more interpretable mutation process than pure scalar ratcheting. CausalEvolve~\citep{chen_causalevolve_2026} extends open-ended discovery with a causal scratchpad that traces how earlier modifications shape later candidates, and Xia et al.~\citep{xia_ai_2026} apply the self-improvement template to the discovery of new LLM-RL algorithms. Open-source projects including autoevolve~\citep{noauthor_mrtsepaautoevolve_2026}, self-improving-agent~\citep{noauthor_peterskoettself-improving-agent_2026}, and self-improving-coding-agent~\citep{noauthor_maximerobeynsself_improving_coding_agent_2026} further illustrate the breadth of self-modification patterns being explored in this part of the ecosystem.

\subsubsection{Multi-Agent Collaborative Research}\label{app:survey:multi-agent}

A fourth branch of the auto-research ecosystem distributes the research process across multiple coordinated agents rather than executing it through a single agent operating sequentially. CORAL~\citep{noauthor_coral_2026} formulates multi-agent research as an evolutionary process for open-ended discovery, with distinct agents filling separated roles in candidate generation and evaluation, and with task improvements emerging from the interaction between roles rather than from a single agent's iteration. The architecture is notable for placing evaluator separation as an explicit design commitment, even though the evaluators themselves remain within the same model family as the producers.

ClawTeam~\citep{noauthor_hkudsclawteam_2026} approaches the multi-agent setting from an orthogonal direction. Rather than separating roles, it runs parallel agent instances across multiple GPU directions in a swarm-style configuration, broadening exploration by sampling more diverse research paths simultaneously before aggregating results through ranked selection. ARIS~\citep{noauthor_wanshuiyinauto-claude-code-research--sleep_2026} moves in yet another direction along several axes at once: it draws reviewer agents from different foundation model families (Claude, Codex, and GPT-style backbones), so that producer and evaluator no longer share training distribution and prompting regime; it organizes research capabilities as composable skill modules invoked on demand rather than as a monolithic pipeline; and it adopts a lightweight, loosely coupled architecture in which the main loop, reviewers, and skill modules communicate through decoupled interfaces and run asynchronously with modest hardware requirements. Shen et al.~\citep{shen_empirical_2026} contribute the most direct empirical examination of these design choices, comparing several multi-agent collaboration patterns on automated research tasks. Their results suggest that the marginal benefit of adding agents depends substantially on whether the collaboration pattern actually changes the information considered at decision points, rather than only the volume of generated candidates.

Several adjacent systems explore narrower multi-agent configurations. ML-Agent~\citep{noauthor_masworksml-agent_2026} targets machine-learning research workflows with a multi-agent decomposition of the pipeline. hyperspaceai/agi~\citep{noauthor_hyperspaceaiagi_2026} proposes a distributed multi-agent architecture for general research automation, and ABSTRAL~\citep{noauthor_abstral_2026} formulates the multi-agent system's topology itself as the object of iterative refinement.

\subsubsection{Evaluation for Auto-Research Loops}\label{app:survey:ecosystem:eval}

A complementary line of work develops evaluation infrastructure that probes auto-research loops along the dimensions where they are most likely to fail. FML-bench~\citep{Zou2026-fk} holds the ML research task fixed and compares agent search strategies, isolating which strategy choices drive performance and supplying a controlled testbed for search-dynamics analysis. SciIntegrity-Bench~\citep{yang2026sciintegrity} scores AI scientist systems on faithful citation, honest result reporting, and reliable claim attribution, targeting the integrity of publication-shaped outputs at termination. MedProbeBench~\citep{liu_medprobebench_2026} extends evaluation to a demanding domain regime by assessing whether agents can synthesize and reconcile heterogeneous evidence sources at the level of expert clinical guideline development, operationalizing a domain-level integration standard. Together these benchmarks probe complementary properties of the auto-research loop: search dynamics during execution, artifact integrity at termination, and conformance to domain-level integration standards.

\subsection{Substrates and Capabilities}\label{app:survey:substrates}

The systems in Appendix~\ref{app:survey:ecosystem} compose research loops out of agent capabilities and persistent state. Two adjacent research programs supply much of the underlying machinery: skill systems study how agents acquire, organize, and invoke reusable procedures, and memory systems study how agents accumulate and retrieve information across episodes. These programs are not themselves auto-research systems, which means they do not by themselves close a research-execution loop. However, they increasingly furnish the substrate on which auto-research loops operate, and a survey of the landscape would be incomplete without them.

\subsubsection{Skill Systems}\label{app:survey:skill}
Skill systems address how agents represent, acquire, and deploy reusable procedural knowledge. The recent survey~\citep{noauthor_agent_2026-1} organizes the field along three axes: architecture (how skills are represented and composed), acquisition (how skills are learned, discovered, or distilled from experience), and security (how skill invocation can be constrained). The framing makes clear that the field is no longer about isolated tool use but about treating procedural capability itself as a structured object that agents can manipulate, share, and improve.

Several systems illustrate the directions surveyed in that framing. Memento-skills~\citep{noauthor_memento-skills_2026} pushes the agent-designs-agent pattern to the skill layer, with agents generating and refining the skills of other agents in a meta-design loop. Seagent~\citep{noauthor_seagent_2025} formulates a self-evolving computer-use agent that acquires skills directly from its own interaction trajectories rather than from a fixed skill library, illustrating how experience-driven skill formation can operate without an externally curated taxonomy.

A first thread of skill-systems work concerns acquisition and evolution. EvoSkill~\citep{noauthor_evoskill_2026} discovers skills through evolutionary search across multi-agent configurations. XSkill~\citep{noauthor_xskill_2026} supports continual learning of skills across multimodal agent tasks, and AutoSkill~\citep{noauthor_autoskill_2026} together with Trace2skill~\citep{noauthor_trace2skill_2026} target lifelong skill self-evolution and trajectory-to-skill distillation respectively. CycleQD~\citep{noauthor_agent_2024} approaches skill acquisition through quality-diversity optimization, and SkillRL~\citep{noauthor_skillrl_2026} evolves agents via recursive skill-augmented reinforcement learning.

A second thread concerns how acquired skills are organized and made retrievable at ecosystem scale. SkillFlow~\citep{noauthor_skillflow_2025} proposes scalable skill retrieval as a system component, SkillNet~\citep{noauthor_skillnet_2026} treats skill creation, evaluation, and connection as a unified workflow, the AI-Research-Skills project~\citep{noauthor_orchestra-researchai-research-skills_2026} assembles a curated skill library targeted at research workflows, and Li et al.~\citep{noauthor_organizing_2026} formulate orchestration and benchmarking of agent skills at ecosystem scale. 

Complementary work probes the limits of skill-based architectures: SWE-Skills-Bench~\citep{noauthor_swe-skills-bench_2026} evaluates whether reusable skills actually help in real-world software engineering, SkillsBench~\citep{noauthor_skillsbench_2026} benchmarks how well agent skills transfer across diverse task settings, Li~\citep{noauthor_when_2026} examines when single-agent-with-skills configurations can replace multi-agent systems and when they fail, and TARSE~\citep{noauthor_tarse_2026} studies test-time skill retrieval as an alternative to fixed skill libraries in reasoning agents.

\subsubsection{Memory Systems}\label{app:survey:memory}

Memory systems address how agents accumulate and retrieve information across episodes, supplying continuity that single-shot agent calls cannot provide. MuSEAgent~\citep{wang_museagent_2026} formulates a multimodal reasoning agent whose state persists across interactions through stateful experiences, with explicit mechanisms for encoding, retrieving, and updating episodic information as the agent operates over multimodal inputs. Omni-SimpleMem~\citep{noauthor_omni-simplemem_2026} takes an unusual inverse approach: rather than designing memory directly, it uses an autoresearch-style loop to discover memory architectures for lifelong multimodal agents, treating memory design itself as the object of autonomous experimentation. EvolveMem~\citep{liu2026evolvemem} pursues the same direction in a self-evolving variant, in which the memory architecture is iteratively rewritten by the agent itself during operation rather than fixed at design time. The three systems together illustrate the range of design points being explored, from memory as a built-in architectural component to memory as an emergent product of automated search.

Memory Intelligence Agent~\citep{noauthor_memory_2026} contributes a further framing of memory as a first-class agent capability rather than as a service of the underlying model. Memory considerations also appear, implicitly or explicitly, throughout the skill-systems work discussed in Appendix~\ref{app:survey:skill}: Seagent~\citep{noauthor_seagent_2025} acquires skills from interaction trajectories that function as a form of episodic memory, and Trace2skill~\citep{noauthor_trace2skill_2026} distills trajectory-local lessons into transferable skills, blurring the line between memory accumulation and skill formation. The boundary between these two systems is becoming blurred, as both are emerging as important components of higher-level agent capabilities that support auto-research.

\subsection{Domain Applications and Boundary Systems}\label{app:survey:domains}

This section turns from system designs to the domains in which auto-research is being deployed. Rather than re-categorizing systems already discussed in previous Appendices, we focus here on systems whose primary contribution lies in adapting auto-research to a particular scientific domain, where domain-specific data, evaluation standards, and stakes shape the design. The final subsection then turns outward to systems that fall outside the LLM-workflow auto-research paradigm but that are frequently raised as adjacent reference points.

\subsubsection{Biomedical and Medical Applications}\label{app:survey:biomed}

Biomedical and medical applications constitute the largest concentration of domain-adapted auto-research work in the current literature, reflecting both the volume of high-quality structured data and the particularly demanding evaluation standards of clinical and biological research. Wu et al.~\citep{wu_towards_2026} present a publication-shaped pipeline specialized for medical research, with the loop adapted to work over clinical questions and biomedical evidence: literature retrieval is grounded in PubMed-scale corpora, hypothesis generation is conditioned on clinical task formulations, and output drafting follows medical-research conventions. The system illustrates the tightening that occurs when general auto-research scaffolding is applied to a domain where evaluators are well-defined and stakes are explicit.

MedOpenClaw~\citep{shen_medopenclaw_2026} pushes the design in a different direction. Rather than operating over curated datasets, the system reasons over uncurated full medical imaging studies---the kind of data clinicians actually encounter---and supplies an auditable reasoning trace through which its decisions can be inspected. The two design commitments together make MedOpenClaw a notable counterpoint to benchmark-tuned medical agents: input realism and reasoning transparency are treated as architectural requirements rather than as post-hoc explainability layers.

Several adjacent systems explore narrower clinical and biological problems. Fan et al.~\citep{fan_evolving_2026} extend medical imaging agents with experience-driven self-skill discovery, allowing the agent to accumulate domain-specific skills from execution trajectories rather than from a fixed skill library. Rhizome OS-1~\citep{noauthor_rhizome_2026} formulates a semi-autonomous operating system for small-molecule drug discovery, coordinating multiple computational and laboratory operations under a unified agent interface. ProtRLSearch~\citep{noauthor_protrlsearch_2026} introduces a multi-round multimodal protein search agent trained via reinforcement learning, targeting structure-aware protein retrieval. MedMASLab~\citep{noauthor_medmaslab_2026} provides a unified orchestration framework for benchmarking multimodal medical multi-agent systems, supplying the evaluation infrastructure that the broader medical auto-research ecosystem increasingly relies on.

\subsubsection{Physics, Chemistry, Mathematics, and Other Sciences}\label{app:survey:other-sci}

Beyond biomedical research, auto-research has begun to take root in the physical sciences, robotics, and adjacent domains, although the overall volume of work remains smaller than in medicine. TRACE~\citep{liu_trace_2026} formulates a multi-agent system for autonomous physical reasoning in seismology, with specialized agents handling waveform interpretation, source-mechanism inference, and seismic-event reasoning in coordination. The design illustrates how physical-science domains push auto-research toward representations of domain-specific physical knowledge that general scaffolding does not supply. Dr.Sai~\citep{he_drsai_2026} approaches a different physics problem: real-world physics analysis at the BESIII collider, an experimental high-energy physics setting with mature analysis conventions and well-defined evaluation standards. The system demonstrates that agentic AI can interface with established experimental-physics workflows, although in such settings the human physicist remains tightly coupled to the analysis loop.

Alexander et al.~\citep{alexander_autonomous_2026} extend the physical sciences direction toward theory induction, formulating autonomous discovery of particle physics theories from experimental data. The work sits at the intersection of symbolic search and LLM-assisted reasoning, illustrating one of the most ambitious targets for autonomous scientific discovery in the physical sciences. Xu et al.~\citep{xu2026beyond} extend autonomous discovery into cosmology, moving the agent from an assistant role toward hypothesis formation and analysis over cosmological data. STRIDE~\citep{Su2026-jw} pursues a related target in symbolic equation discovery, introducing a self-reflective agent framework that combines data-aware generation with mixed-fitting evaluation to improve the reliability of recovered equations. Nomad~\citep{jia_nomad_2026} pursues a more general autonomous exploration and discovery framework, designed to operate across domains rather than within a single scientific specialty. Somasekharan et al.~\citep{somasekharan2026ai} extend autonomous discovery into computational fluid dynamics, using physics-aware agents to explore open-ended CFD problems beyond predefined task templates. Xu et al.~\citep{xu2026agentic} approach the mathematical-sciences end of the spectrum with agentic MIP research, accelerating the generation of constraint handlers for mixed-integer programming through agent-driven search.

A materials-science cluster pursues autonomous discovery through direct integration with experimental and computational characterization. Qumus~\citep{Shi2026-de} realizes an embodied AI experimentalist that integrates hypothesis generation, robotic experimentation, and analysis within a single mini-laboratory for van der Waals heterostructures, while Lee et al.~\citep{Lee2026-be} coordinate multiple instruments in real time to autonomously discover phase-change memory materials in the Mn-Sb-Te ternary system. Cobelli et al.~\citep{cobelli2026agentic} target the upstream representation problem in materials discovery, using an autoresearch-style loop to design compositional descriptors that downstream models can exploit.

Adjacent applications extend into robotics and the social sciences. Khandelwal and Gupta~\citep{khandelwal_agent-driven_2026} formulate agent-driven autonomous reinforcement learning research for quadruped locomotion, with an agent iteratively proposing policy improvements that are evaluated in simulated and physical environments. LLM Agents as Social Scientists~\citep{noauthor_llm_2026} extends the auto-research template into the social sciences, supplying a human-AI collaborative platform for social-science research automation. The two illustrate how the auto-research paradigm is being adapted both toward physically grounded experimentation and toward domains where the unit of investigation is social rather than natural.

\subsubsection{Boundary Systems Beyond Closure-for-Autonomy}\label{app:survey:boundary}

A separate family of systems is frequently raised in connection with auto-research but lies outside the LLM-workflow paradigm surveyed in this appendix. What distinguishes these systems is that an external operation, such as formal verification, physical experimentation, expert sign-off, or executable evaluation, which is constitutive of their architecture rather than supplementary to it. We catalog them here as a complement to the discussion of alternative views in Section~\ref{sec:alternative-views}, which addresses several of these systems in the context of the paper's scope.

A first cluster grounds the system in formal or physical execution. The semi-autonomous formalization of the Vlasov-Maxwell-Landau equilibrium~\citep{ilin_semi-autonomous_2026} couples LLM-generated proof attempts to a Lean 4 verifier together with in-loop mathematician supervision, so that the system's notion of a valid step is fixed by formal proof correctness. Coscientist~\citep{boiko_autonomous_2023} integrates LLM-driven chemical research with cloud-laboratory execution, with physical experimentation incorporated directly into the system's operation. Latent-Y~\citep{team_latent-y_2026} extends this pattern to de-novo drug design, with antibody-design campaigns paired with wet-laboratory assays as integral validation steps in each design cycle. Robin demonstrates the same pattern in autonomous therapeutic discovery~\citep{Ghareeb2026-sk}: multi-agent literature search and data analysis are coupled with in-vitro assays as a constitutive validation step, and the system identifies ripasudil and KL001 as candidate treatments for dry age-related macular degeneration, with efficacy confirmed in the laboratory before the loop accepts the result. Although the multi-agent ideation and analysis layer resembles closure-for-autonomy systems on the surface, the lab-in-the-loop structure relocates the validity condition outside the producer's inductive boundary. CRISPR-GPT~\citep{noauthor_crispr-gpt_2024} agentically  automates gene-editing experiments while making expert validation an architectural step in the experiment-design pipeline. These four laboratory-grounded systems together demonstrate that auto-research-adjacent architectures in the life sciences typically incorporate physical or expert validation as constitutive components rather than as optional add-ons.

A second cluster grounds the system in executable evaluation. FunSearch~\citep{romera-paredes_mathematical_2023} filters LLM-generated mathematical and algorithmic programs through executable evaluators, ensuring that candidate outputs are vetted by formal program execution before being retained. AlphaEvolve~\citep{novikov_alphaevolve_2025} generalizes this approach into a coding agent for scientific and algorithmic discovery, with executable algorithmic evaluation serving as the validity condition. ERA~\citep{Aygun2026-qy} extends the same paradigm to scientific software generation, using tree search over executable quality metrics on public leaderboards and external baselines across single-cell analysis, COVID-19 hospitalization forecasting, and additional scientific software tasks. A further pair of systems sits adjacent to auto-research in a different way. Google's AI Co-Scientist~\citep{gottweis_towards_2025} is primarily a scientist-guided hypothesis-generation system whose outputs require downstream experimental validation, with its \emph{Nature} 2026 deployment reporting in-vitro confirmation of drug repurposing candidates for acute myeloid leukemia, with the scientist positioned at the center of the system's design. OpenAI's Deep Research~\citep{openai_deep_research_2025} is a source-grounded research-synthesis product rather than an autonomous discovery loop. The broader deep-research category, including its verification-centric and meta-evaluation lines, is surveyed and developed in~\citep{huang_deep_2025, zhu_marco_2026, team_mirothinker-17_2026, Wang2026-ks}.

\subsection{Concurrent Diagnostic Positions}\label{app:survey:concurrent}

Alongside the rapidly growing system literature surveyed in the preceding sections, a complementary body of work asks adjacent diagnostic questions about whether the outputs of these systems deserve the standing their surface form suggests. Five works are particularly relevant for situating the auto-research landscape and warrant focused treatment. Ríos-García et al.~\citep{rios-garcia_ai_2026} evaluate LLM-based scientific agents across eight domains and more than twenty-five thousand agent runs, combining performance decomposition with behavioral analysis of epistemological traces. Their central finding is that current systems can execute scientific workflows without reliably exhibiting the reasoning patterns that scientific practice typically requires: substantive evidence consideration, refutation-driven belief revision, and convergent multi-test reasoning often fail to appear in agent traces even when the workflow completes to publication-shaped form. The diagnosis locates the gap at the level of how individual runs handle evidence, contradiction, and convergence.

Two related works use Karpathy's \texttt{autoresearch} template as a controlled testbed for diagnostic study. Li~\citep{li_auto_2026} conducts a convergence analysis across ten thousand experiments and finds that what is nominally architecture search can reduce to hyperparameter tuning within a narrow design region, with the loop's internal scoring driving the agent toward portions of the design space that the score most readily rewards. The result documents an empirical narrowing effect that follows from optimizing nominally broad search against a single internal signal. He et al.~\citep{he_autoresearch_2026} take a different angle from the same starting point, framing the rise of auto-research as a methodological moment in which the human role shifts from experimenter to research director: as loops execute more of the research workflow autonomously, the practitioner's task increasingly becomes the specification, oversight, and stewardship of loops rather than the direct conduct of individual experiments.

Zhang et al.~\citep{Zhang2026-fg} provide the most direct empirical test of whether artifact production aligns with scientific substance.Letting frontier coding agents carry out the full research loop across thirteen computer-science seeds, they produce 117 agent-generated papers. Under manuscript-only automated review the strongest system scores at the weighted-average level of ICLR 2025 submissions; under artifact-aware review, in which evaluators inspect the workspace alongside the manuscript,scores drop sharply, with manual auditing identifying experimental rigor as the dominant failure mode and none of the 117 papers meeting the acceptance bar of a top-tier venue. This empirical pattern aligns with the diagnostic framework of the present paper: the gap between manuscript-only and artifact-aware review is what the L2 collapse predicts at the level of loop architecture, and the dominance of experimental rigor among failure modes is consistent with the L1 collapse acting upstream of validation.

Messeri and Crockett~\citep{messeri_artificial_2024} broaden the diagnostic frame beyond auto-research specifically, arguing that AI tools can produce illusions of understanding in scientific research when they bypass the distributed processes through which scientific knowledge is normally critiqued, integrated, and revised. An accompanying \emph{Nature} editorial published alongside the recent Co-Scientist and Robin systems further argues that AI should empower scientific work but not replace human researchers~\citep{noauthor_2026-lw}, placing the question of autonomous research in an explicitly normative frame. Although their account is not specific to LLM-workflow auto-research, it identifies a gap between fluent artifact production and durable knowledge production that becomes increasingly visible as auto-research systems generate more publication-shaped outputs. 

A fifth complementary position takes the form of a contemporaneous life-cycle survey rather than a focused diagnosis. Kong et al.~\citep{Kong2026-as} organize AI-assisted research across creation, writing, validation, and dissemination, and identify a stage-dependent boundary between reliable assistance and unreliable autonomy, with human-governed collaboration framed as the most credible deployment mode. This stage-level reading and our loop-level reading are complementary rather than competing: where the former asks which research stages support autonomy, the latter asks how a loop's internal control structure decides what counts as progress, validity, and standing.

The five positions sketched here approach auto-research from distinct analytic angles, but none analyzes the field through the lens of scientific closure. Our work is complementary to these positions and fills this gap.

%% file: Appendix/Appendix_audits.tex
\section{Audit Methodology and Per-System Coding}\label{app:audit}

\subsection{From Survey to Audit}\label{app:audit:overview}

Appendix~\ref{app:survey} surveys the landscape of LLM-workflow auto-research at descriptive breadth, covering over one hundred works across systems, evaluation infrastructure, and adjacent diagnostic positions. This appendix narrows from that landscape to analytic depth: it documents the coding scheme used to assign the labels reported in Table~\ref{tab:collapse-audit} of Section~\ref{sec:empirical-audit}, and provides the per-system evidence supporting each assignment. The 21 systems in the audit pool are a subset of the broader corpus, selected because their loop structure is publicly documented in sufficient detail to support coding against a fixed scheme. The purpose is not to recategorize the systems already described in Appendix~\ref{app:survey}, but to make the basis for the audit's quantitative claims inspectable and reproducible. Coding draws only on public artifacts, such as papers, repositories, documentation, prompts, and reported workflows.

\subsection{Coding Scheme}\label{app:audit:scheme}

Each of the 21 systems is coded on three dimensions corresponding to the three-level collapse introduced in Section~\ref{sec:three-collapses}. Labels are assigned based on the role a design feature plays in the loop's control structure. That is, whether and how it actually gates the loop's selection, validation, or termination behavior. Each dimension admits three possible labels:

\begin{itemize}
\item \textbf{Strong}: the dimension is not addressed; the auto-research system's loop behavior is governed entirely by internal signals.
\item \textbf{Weak}: the dimension is addressed by mechanisms that simulate external involvement from within the system's inductive boundary.
\item \textbf{Mitigated}: the dimension is addressed by a genuinely external operation that lies outside the system's inductive boundary and is integrated as a binding condition of loop validity.
\end{itemize}

The separation between Weak and Mitigated is the central operational distinction in the audit: Weak captures systems whose external-looking checks are simulated from within the system's inductive boundary, whereas Mitigated requires a genuinely external operation integrated into the loop as a binding condition of validity. The absence of Mitigated in Table~\ref{tab:collapse-audit} is a feature of the paradigm under study, not a gap in coverage.

\subsection{Per-System Audit Trace}\label{app:audit:trace}

This appendix records the per-system evidence supporting the labels reported in Section~\ref{sec:empirical-audit}. For each of the 21 audited systems, Table~\ref{tab:audit-pool} reports the objective signal, validator, and output pathway based on public documentation and literature synthesis, together with collapse codes for the three dimensions defined in Section~\ref{sec:empirical-audit} and Appendix~\ref{app:audit:scheme}. The table is intended as an audit trace under those coding rules, not as a comprehensive system catalog. The 21 systems were selected because each has sufficient public documentation (papers, repositories, or both) to support coding without reliance on undocumented internal practices. No system in the audit pool receives a mitigated label on any dimension, under the rules in Appendix~\ref{app:audit:scheme}.


{\setlength{\LTcapwidth}{\linewidth}\setlength{\tabcolsep}{3pt}\renewcommand{\arraystretch}{1.05}\begin{longtable}{>{\footnotesize\raggedright\arraybackslash}p{0.16\linewidth}                  >{\footnotesize\centering\arraybackslash}p{0.025\linewidth}                  >{\footnotesize\centering\arraybackslash}p{0.025\linewidth}                  >{\footnotesize\centering\arraybackslash}p{0.025\linewidth}                  >{\footnotesize\raggedright\arraybackslash}p{0.22\linewidth}                  >{\footnotesize\raggedright\arraybackslash}p{0.22\linewidth}                  >{\footnotesize\raggedright\arraybackslash}p{0.22\linewidth}}\caption{Per-system audit. Each row records the method's objective signal, validator, and output pathway, with collapse codes for the three dimensions (S = strong collapse; W = weak collapse).}\label{tab:audit-pool}\\\toprule\textbf{System} & \textbf{L1} & \textbf{L2} & \textbf{L3} & \textbf{Objective signal} & \textbf{Validator} & \textbf{Output pathway}\\\midrule\endfirsthead\toprule\textbf{System} & \textbf{L1} & \textbf{L2} & \textbf{L3} & \textbf{Objective signal} & \textbf{Validator} & \textbf{Output pathway}\\\midrule\endhead
Karpathy autoresearch\citep{andrej_karpathyautoresearch_2026} & S & S & S & Benchmark or experiment score & Internal inspection plus benchmark feedback & Improved code solution\\uditgoenka-autoresearch\citep{noauthor_uditgoenkaautoresearch_2026} & S & S & S & Internal task metric or experiment result & Internal review and execution feedback & Improved repository artifact \\goal-md\citep{noauthor_jmilinovichgoal-md_2026} & S & S & S & Agent-constructed fitness criterion & Internal execution and metric feedback & Goal-conditioned code or task artifact \\autoresearch-anything\citep{noauthor_zkarimi22autoresearch-anything_2026} & S & S & S & Measurable task score & Internal execution and scoring & Task-specific repository artifact \\ADAS\citep{noauthor_shengranhuadas_2026} & S & S & S & Candidate architecture performance & Benchmark-mediated evaluation & Improved agent architecture \\AIDE~\citep{noauthor_wecoaiaideml_2026} & S & S & S & User-defined scalar metric(e.g., MLE-Bench, Kaggle leaderboard)  & Internal execution feedback within the search regime & Improved code solution \\GEPA\citep{noauthor_gepa-aigepa_2026} & W & S & S & Ranked task performance after reflection & Internal reflection and task feedback & Optimized prompt, code, or text \\HGM\citep{noauthor_metauto-aihgm_2026} & S & S & S & SWE-bench performance & Internal execution and self-review & Improved agent or codebase \\Bilevel Autoresearch\citep{qu_bilevel_2026} & S & S & S & Task score plus meta improvement rate & Internal benchmark and execution feedback at both levels & Improved loop plus task artifact \\ASI-Evolve\citep{xu_asi_evolve_2026} & W & S & S & Aggregated system-level performance gain & Experimental benchmark feedback and internal analyzer & Discovered architectures, data, or algorithms \\Omni-SimpleMem\citep{noauthor_omni-simplemem_2026} & S & S & S & Benchmark improvement on memory tasks & Internal experiment execution and benchmark feedback & Memory framework and design changes \\Claudini\citep{panfilov_claudini_2026} & S & W & S & Attack success and transfer performance & Held-out adversarial evaluation within benchmark regime & Released attack algorithms and evaluation code \\CORAL\citep{noauthor_coral_2026} & S & W & S & Task improvement rates and evaluator scores & Evaluator separation and task feedback inside system & Improved mathematical, algorithmic, or systems solutions \\SakanaAI/AI-Scientist\citep{noauthor_sakanaaiai-scientist_2026,lu_towards_2026} & S & W & W & Automated reviewer score & Automated review plus workshop peer review & Generated paper plus workshop-tested output (Submit to ICLR 2025 Workshop) \\AI-Scientist-v2\citep{noauthor_sakanaaiai-scientist-v2_2026} & S & S & S & Internal reviewer or evaluation score & Internal automated evaluation & Generated research artifact \\AutoResearchClaw\citep{noauthor_aiming-labautoresearchclaw_2026} & S & W & S & Paper-generation objective and internal selection & Citation verification plus internal review & Publication-shaped output \\Agent Laboratory\citep{noauthor_samuelschmidgallagentlaboratory_2026,schmidgall_agentrxiv_2025} & S & S & S & Phase-internal completion plus paper-quality signals & Internal review across same-family agents & Generated paper plus AgentRxiv-mediated access but not external correction \\CycleResearcher\citep{wengcycleresearcher} & S & S & S & Averaged reviewer score used as SimPO reward & Same-family reward model trained on ICLR-2024 review data & Watermarked LaTeX paper; submission to real peer review prohibited \\AI-Researcher \citep{noauthor_hkudsai-researcher_2026} & W & W & W & Aggregated objectives across multiple tasks & Cross-model review and code review & Novix.science production-facing deployment \\ARIS\citep{noauthor_wanshuiyinauto-claude-code-research--sleep_2026} & S & W & S & Internal aggregated score & Cross-model review across Claude, Codex, and GPT-style systems & Terminal research artifact \\ClawTeam\citep{noauthor_hkudsclawteam_2026} & W & S & S & Aggregated ranked result after parallel search & Internal validation & Terminal artifact \\\bottomrule\end{longtable}}\subsection{Coding Limitations}\label{app:audit:limits}

The coding is based on public artifacts and undocumented internal practices cannot alter labels. The Strong/Weak distinction is intentionally coarse and identifies structural patterns in loop control, not overall system quality, novelty, or usefulness. The audit identifies co-occurrence between closure-for-autonomy and the three collapse dimensions but does not establish that one dimension causally produces another.